# Ni-O-Ag catalyst enables 103-m$^2$ artificial photosynthesis with >16% solar-to-chemical energy conversion efficiency

Yaguang Li[1, 4, *], Fanqi Meng[2, 4], Qixuan Wu[1, 4], Dachao Yuan[3, 4], Haixiao Wang[1], Bang Liu[1], Junwei Wang[1], Xingyuan San[1], Lin Gu[2],Shufang Wang[1], Qingbo Meng[2, *]

[1]Research Center for Solar Driven Carbon Neutrality, The College of Physics Science and Technology, Institute of Life Science and Green Development, Hebei University, Baoding, 071002, China.
[2]Beijing National Laboratory for Condensed Matter Physics, Institute of Physics, Chinese Academy of Sciences, Beijing, 100190, China.
[3]College of Mechanical and Electrical Engineering, Hebei Agricultural University, Baoding 071001, China.
[4]These authors contributed equally to this work.

Correspondence and requests for materials should be addressed to Y. Li. (email: liyaguang@hbu.edu.cn) or to Q. Meng. (email: qbmeng@iphy.ac.cn).

## Abstract

Herein, NiO nanosheets supported with Ag single atoms (2D $Ni_1Ag_{0.02}O_1$) are synthesized for photothermal $CO_2$ hydrogenation to achieve 1065 mmol $g^{-1}$ $h^{-1}$ of CO production rate under 1 sun irradiation, revealing the unparalleled weak sunlight driven reverse water-gas shift reaction (RWGS) activity. This performance is attributed to the coupling effect of Ag-O-Ni sites to enhance the hydrogenation of $CO_2$ and weaken the CO adsorption, resulting in 1434 mmol $g^{-1}$ $h^{-1}$ of CO yield at 300 ℃, surpassing any low-temperature RWGS performances ever reported. Building on this, we integrated the 2D $Ni_1Ag_{0.02}O_1$ supported photothermal RWGS with commercial photovoltaic electrolytic water splitting, leading to the realization of 103-$m^2$ scale artificial photosynthesis system ($CO_2+H_2O \rightarrow CO+H_2O$) with a daily CO yield of 18.70 $m^3$, a photochemical energy conversion efficiency of >16%, over 90% $H_2$ ultilazation efficiency, outperforming other types of artificial photosynthesis. The results of this research chart a promising course for designing practical, natural sunlight-driven artificial photosynthesis systems and highly efficient platinum-free $CO_2$ hydrogenation catalysts. This work is a significant step towards harnessing solar energy more efficiently and sustainably, opening exciting possibilities for future research and development in this area.

## Introduction

Artificial photosynthesis, which uses solar energy to convert $CO_2$ into chemicals and fuels, is emerging as a promising path towards carbon neutrality.[1,2] Given that carbon monoxide (CO) is a vital precursor for many valuable fuels and chemicals in various industries,[3,4] numerous artificial photosynthetic systems have been developed for solar driven CO generation by using $CO_2$ and $H_2O$ ($CO_2+H_2O \rightarrow CO+O_2$).[5,6] With the developement of various efficient catalysts, such as metals,[7-9] metal compounds,[10-12] molecular complexes,[13,14] the size and solar to chemical energy efficiency (STC) of artificial photosynthetic systems have made great progress (for example, the device with 6.5 $m^2$ size, 3.8% STC[15] and the device with 1 $cm^2$ size and 12.7% STC[16]), revealing the dawn of applicable artificial photosynthesis. However, the state of the art of STC and scale level are still far from meeting practical needs. Therefore, it is one of the holy grails for scientists to synergistically further improve the STC and scale of artificial photosynthesis.

Recognizing that artificial photosynthesis comprises two main processes of water splitting and $CO_2$ hydrogenation,[17,18] a novel artifical photosynthesis paradigm has been suggested: the integration of photovoltaic-electrolytic water splitting and photothermal $CO_2$ hydrogenation (Scheme 1),[19] with ability for large-scale implementation. However, for practical applications, these systems need to operate under ambient conditions. While photovoltaic-electrolytic water splitting can already work under natural sunlight irradiation, the photothermal reverse water-gas shift reaction (RWGS, $CO_2+H_2 \rightarrow CO+H_2O$) necessitates high brightness sunlight irradiation (light intensity > 10 kW $m^{-2}$ = 10 suns) to create high temperature (>450 °C) to maintain operation.[20,21]

Consequently, such systems are unable to function outdoors under typical sunlight conditions using standard photothermal configurations. To address this challenge, in addition to increasing the ambient sunlight driven temperature, the key is to decrease the reaction temperature of RWGS. This has led scientists to develop catalysts based on platinum-group metals that are highly active for RWGS at low temperatures.[22-25] For instance, Ma et al. reported a Pt-$MoO_x$/$Mo_2N$ catalyst that achieves a RWGS CO generation rate of 619.2 mmol $g^{-1}$ $h^{-1}$ at 300 °C.[26] Despite these advances, platinum-group-metal-based catalysts have not yet demonstrated functionality under ambient sunlight and the use of platinum group metals (Pt, Rh, Ru, Pd) significantly increase their cost. Consequently, developing platinum-group-metals free RWGS catalysts with excellent low-temperature activity (<300 °C) and low cost is crucial for realizing efficient ambient sunlight driven photothermal RWGS.

In this study, we designed a catalyst that loaded Ag single atoms on NiO support (2D $Ni_1Ag_{0.02}O_1$) for low-temperature RWGS to show ~~a~~ CO yield of 418.95 mmol $g^{-1}$ $h^{-1}$ and 1434 mmol $g^{-1}$ $h^{-1}$ at 250 °C and 300 °C, respectively, exceeding to any previously reported counterparts. Characterizations and theoretical calculations revealed that the Ag-O-Ni synergistic sites facilitated asymmetric activation of $CO_2$ and the weak adsorption of CO, thus enhancing the RWGS activity. This low-temperature catalytic performance enabled 2D $Ni_1Ag_{0.02}O_1$ to realize efficient weak sunlight driven photothermal RWGS, with a CO production rate of 1065 mmol $g^{-1}$ $h^{-1}$ under 1 sun illumination. Owing to its ability to operate efficiently under weak sunlight irradiation, 2D $Ni_1Ag_{0.02}O_1$ assisted photothermal RWGS could be paired with photovoltaic electrolytic $H_2O$ decomposition to convert $CO_2$ and $H_2O$ into CO and $O_2$ under outdoor sunlight irradiation with a scale of 103 $m^2$ and a STC of >16% throughout the daytime, paving a new benchmark for large-sized artificial photosynthesis.

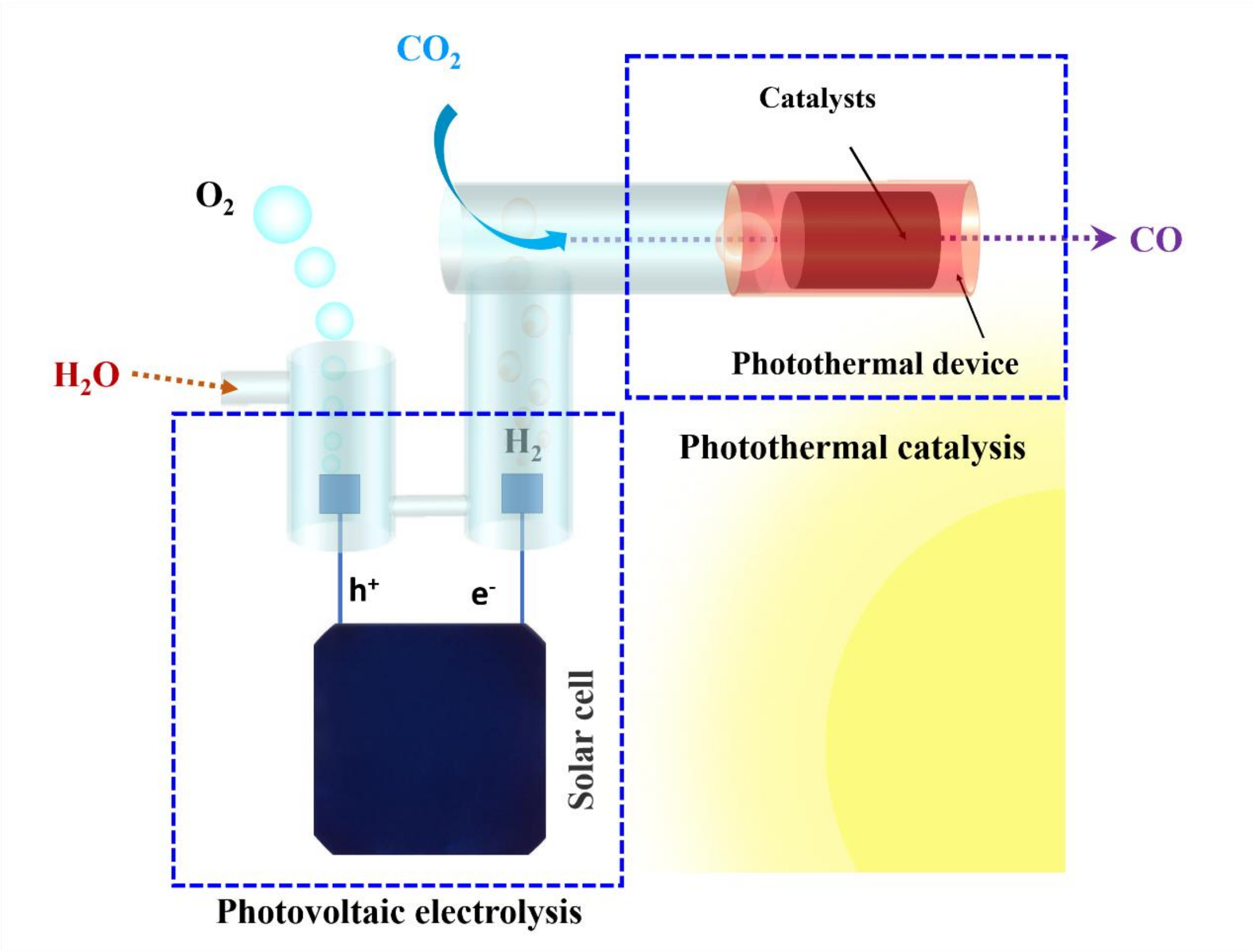

**Scheme 1.** Schematic map of artificial photosynthesis composed by photovoltaic electrolytic water splitting and photothermal $CO_2$ hydrogenation.

## Results and discussions

### Ag single atoms supported on NiO

In a typical synthesis process, we added $Ni(NO_3)_3 \cdot 6H_2O$, $AgNO_3$ and water-soluble starch into the water to form a homogeneous solution. And the solution was frozen with liquid nitrogen and removed the water by lyophilization. Then, an annealing process in the air was applied to form the catalyst (synthesis details are seen in Methods). Fig. 1a, b present the typical scanning electron microscopy (SEM) and transmission electron microscopy (TEM) images of the as-prepared catalyst. The catalyst was grown in two-dimensional morphology and no clear Ag nanoparticles existed in the visual field. Powder X-ray diffraction (XRD) pattern showed that the catalyst had only peaks assigned to NiO rather than Ag based species (Supplementary Fig. 1).[27] The high resolution (HRTEM) image of the catalyst revealed typical NiO(111) crystal planes with the identical lattice spacing of 0.242 nm (Fig. 1c)[28], while X-ray photoelectron spectroscope (XPS) also confirmed the oxidation state of Ni (Supplementary Fig. 2).[29] Further, elemental mapping images demonstrated the homogeneous distribution of O, Ni and Ag, throughout the sample (Fig. 1d), confirming the existence of Ag species. To visualize the Ag species, high angle annular dark-field scanning TEM (HAADF-STEM) was applied and Fig. 1e revealed numerous star-like bright spots on the surface of the sample, which were assigned to Ag single atoms.[30] The atomic ratio of Ag in the sample was 2%. Therefore, we named the catalyst as 2D $Ni_1Ag_{0.02}O_1$. Atomic force microscopy (AFM) confirmed that the thickness of 2D $Ni_1Ag_{0.02}O_1$ was 4 nm, revealing its ultrathin nature (Fig. 1f). We used the X-ray absorption spectroscope (XAS) to characterize the coordination structure of Ag in 2D $Ni_1Ag_{0.02}O_1$. As shown in Fig. 1g, the energy of Ag K-edge in 2D $Ni_1Ag_{0.02}O_1$ was higher than that of Ag foil and lower than that of $Ag_2O$.[31] It indicated that the Ag single atoms in 2D $Ni_1Ag_{0.02}O_1$ slightly oxidation (Supplementary Fig. 3).[32] Fourier transformed EXAFS (FTEXAFS) of 2D $Ni_1Ag_{0.02}O_1$ showed a peak located at 1.6 Å, corresponding to the Ag-O coordination (Fig. 1h).

Moreover, FTEXAFS showed no other peaks for 2D $Ni_1Ag_{0.02}O_1$,[33] confirming the single atomic state of Ag. Computational simulation identified that the Ag-O coordination number of Ag in 2D $Ni_1Ag_{0.02}O_1$ was 1 (Supplementary Table 1).[34] Based on the EXAFS results, we built the atomic structure of 2D $Ni_1Ag_{0.02}O_1$. As shown in Fig. 1i, the Ag single atom was bonded to the lattice oxygen of NiO support, which aligned well with the experimentally obtained FT-EXAFS curve of 2D $Ni_1Ag_{0.02}O_1$ (Fig. 1i).

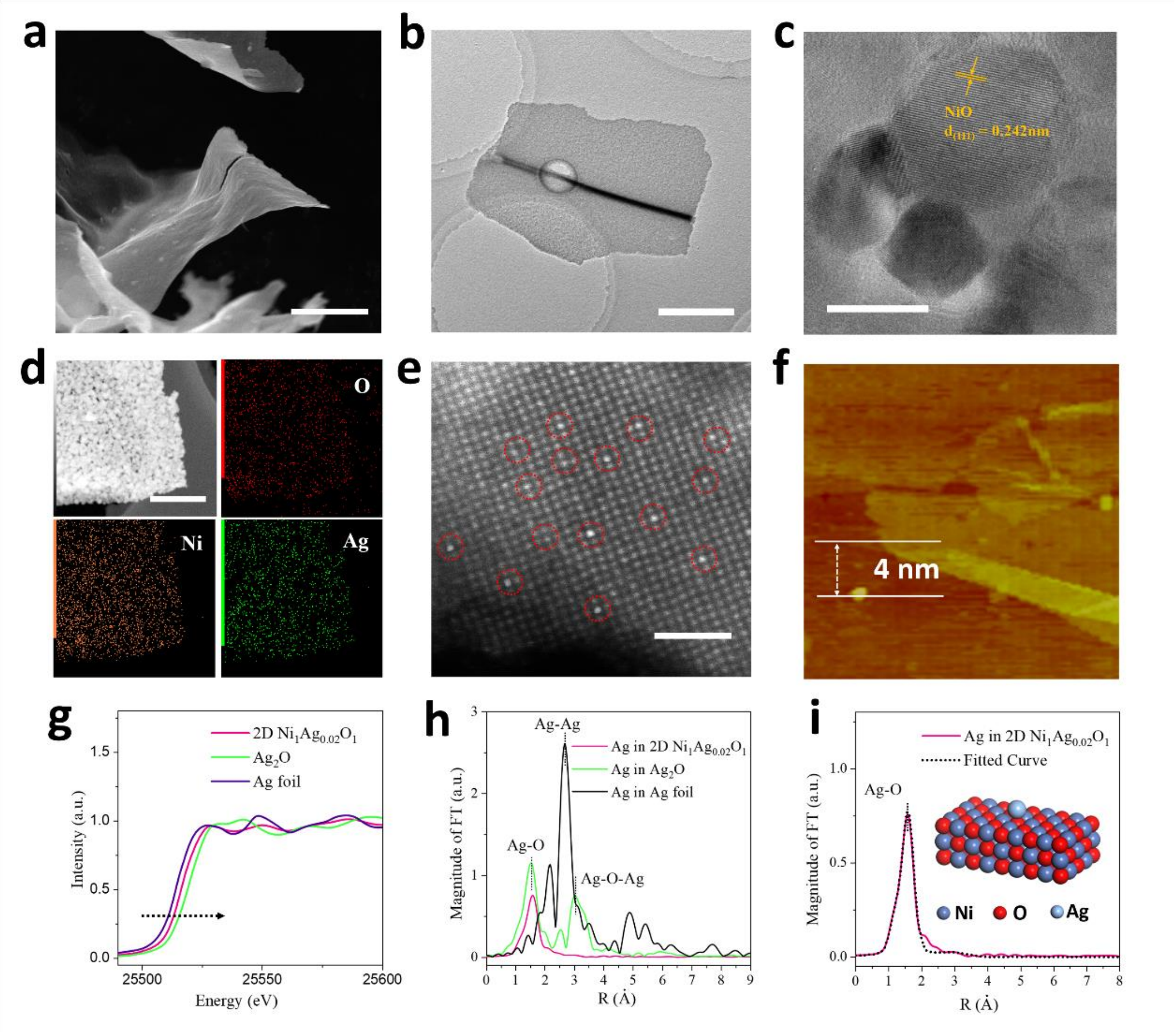

**Fig. 1** The characterization of 2D $Ni_1Ag_{0.02}O_1$. **a** SEM image, **b** TEM image, **c** HRTEM image of 2D $Ni_1Ag_{0.02}O_1$. **d** STEM image and EDS mapping of 2D $Ni_1Ag_{0.02}O_1$. **e** HAADF-STEM image, **f** AFM image of 2D $Ni_1Ag_{0.02}O_1$. **g, h** XANES spectra and FTEXAFS spectra at the Ag K-edge for 2D $Ni_1Ag_{0.02}O_1$, $Ag_2O$, and metallic Ag foil. **i** The FTEXAFS curves of the proposed 2D $Ni_1Ag_{0.02}O_1$ structure (balck line) and the measured 2D $Ni_1Ag_{0.02}O_1$ (purple line). Inset is the proposed model of 2D $Ni_1Ag_{0.02}O_1$ architecture. The scale bars in **a, b, c, d, e** are 2 μm, 1 μm, 5 nm, 500 nm, 1 nm, respectively.

**The RWGS performance of 2D $Ni_1Ag_{0.02}O_1$**

Fig. 2a shows the RWGS performance of 2D $Ni_1Ag_{0.02}O_1$, Ni nanosheets (defined as Ni, Supplementary Fig. 4-6, synthesis details can be found in the Methods sections), Ag nanoparticles (defined as Ag, Supplementary Fig. 7-10). For a fair comparison, we set the activation temperature at a point where the RWGS CO production rate is higher than 1 mmol $g^{-1}$ $h^{-1}$. The RWGS activation temperature of 2D $Ni_1Ag_{0.02}O_1$, Ni was 150 °C, 275 °C, respectively, and Ag remained inactive throughout the entire reaction temperature range (Fig. 2a). Further, the CO production rate of 2D $Ni_1Ag_{0.02}O_1$ reached 1434 mmol $g^{-1}$ $h^{-1}$ at 300 °C with 96.7% CO selectivity (Supplementary Fig. 11), 10.11% $CO_2$ conversion efficiency (Supplementary Fig. 12), exhibiting higher activity and selectivity in comparison with Ni (93.1 mmol $g^{-1}$ $h^{-1}$ of CO production rate and 90% CO selectivity at 300 °C, Supplementary Fig. 13). Fig. 2b and Supplementary Table 2 list the RWGS CO production rate of 2D $Ni_1Ag_{0.02}O_1$ and advanced catalysts at different temperatures. The CO production rate of 2D $Ni_1Ag_{0.02}O_1$ at 300 °C, 250 °C is 1434, 418.95 mmol $g^{-1}$ $h^{-1}$, respectively, not only totally outperformed the best catalysts at corresponding temperatures (619.2 mmol $g^{-1}$ $h^{-1}$ at 300 °C, 277.2 mmol $g^{-1}$ $h^{-1}$ at 250 °C),[26] but also exceeded the catalysts working at higher temperature (350-430 °C, 434.7 mmol $g^{-1}$ $h^{-1}$ at 350 °C).[22] The 2D $Ni_1Ag_{0.02}O_1$ also showed a robust CO production rate of ~870 mmol $g^{-1}$ $h^{-1}$ in a continuous 76-hour RWGS test at 275 °C (Fig. 2c), verifying the excellent stability of 2D $Ni_1Ag_{0.02}O_1$. The specific surface area of 2D $Ni_1Ag_{0.02}O_1$ was 31.5 $m^2$ $g^{-1}$ $h^{-1}$ (Supplementary Fig. 14), ~1.7 times of Ni (17.4 $m^2$ $g^{-1}$, Supplementary Fig. 15). It indicates that the excellent RWGS activity of 2D $Ni_1Ag_{0.02}O_1$ mainly comes from high intrinsic activity of active sites. Fig. 2d shows that the apparent activation energy (Ea) for 2D $Ni_1Ag_{0.02}O_1$ was around 7.55 kJ $mol^{-1}$, which was 1/8 of Ni (~58.66 kJ $mol^{-1}$), implying the great catalytic activity of 2D $Ni_1Ag_{0.02}O_1$.

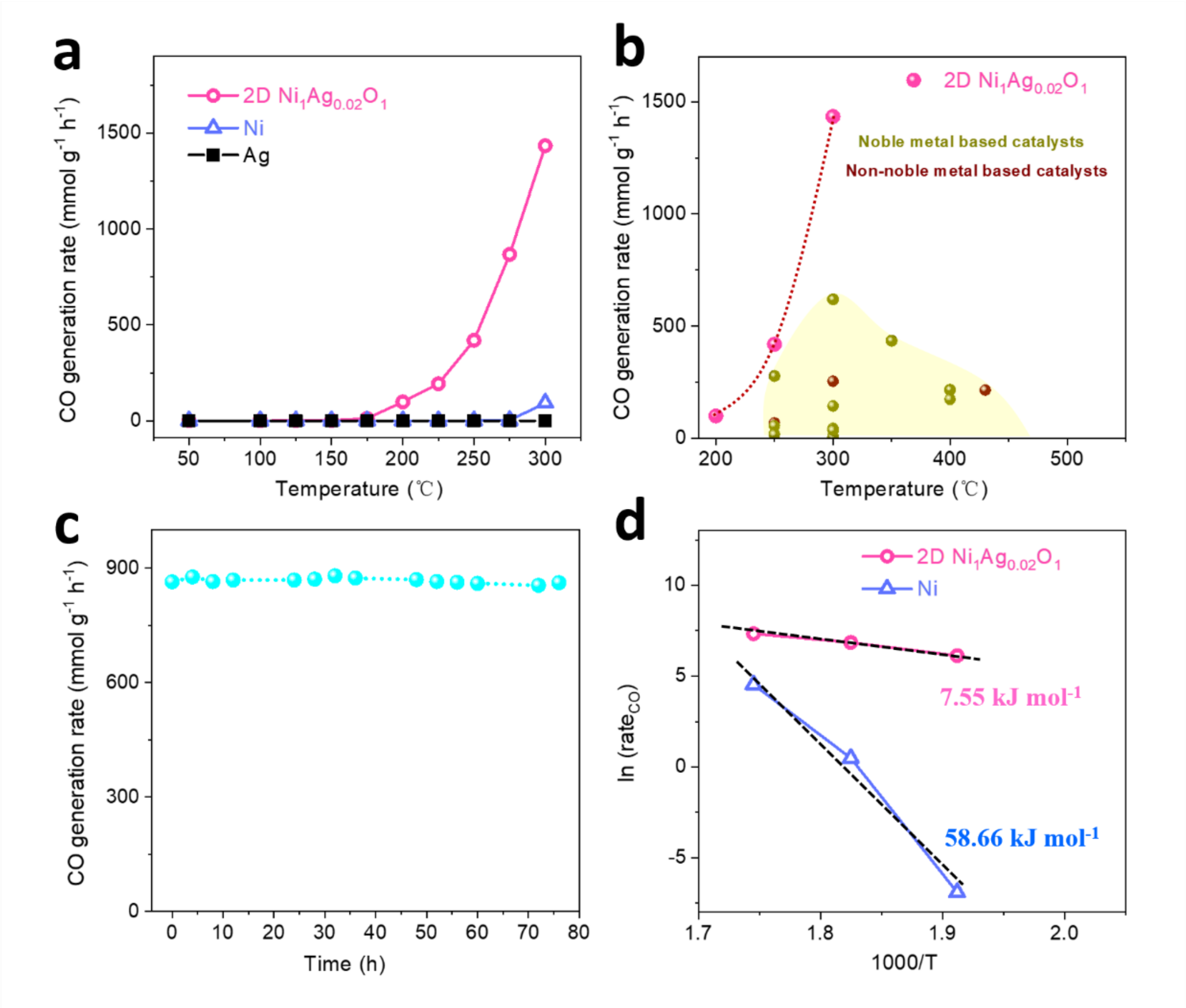

**Fig. 2** RWGS performance and mechanism of 2D $Ni_1Ag_{0.02}O_1$. **a** The CO produced rate

of thermocatalytic RWGS over 2D $Ni_1Ag_{0.02}O_1$, Ni, Ag. **b** The RWGS CO production rate of 2D $Ni_1Ag_{0.02}O_1$ and other advanced catalysts at different temperatures: Pt-$MoO_x/Mo_2N$ (619.2 mmol $g^{-1}$ $h^{-1}$ at 300 °C),[26] $Cu/Mo_2C$ (253.6 mmol $g^{-1}$ $h^{-1}$ at 300 °C),[35] Fe-Pt/$CeO_2$ (142.9 mmol $g^{-1}$ $h^{-1}$ at 300 °C),[22] Pt/$CeO_2$ (43.2 mmol $g^{-1}$ $h^{-1}$ at 300 °C),[36] Na-Rh/$ZrO_2$ (37.4 mmol $g^{-1}$ $h^{-1}$ at 300 °C),[23] PtCo/$TiO_2$ (34.1 mmol $g^{-1}$ $h^{-1}$ at 300 °C),[37] Ru@$MoO_{3-x}$ (7.5 mmol $g^{-1}$ $h^{-1}$ at 300 °C),[24] Rh-$In_2O_{3-x}(OH)_y$ (2.4 mmol $g^{-1}$ $h^{-1}$ at 300 °C),[38] Pt-$MoO_x/Mo_2N$ (277.2 mmol $g^{-1}$ $h^{-1}$ at 250 °C),[26] $Co_2C$ (68.04 mmol $g^{-1}$ $h^{-1}$ at 250 °C),[39] Fe-Pt/$CeO_2$ (54.48 mmol $g^{-1}$ $h^{-1}$ at 250 °C),[22] Ru-Mo-$O_x$ (18.75 mmol $g^{-1}$ $h^{-1}$ at 250 °C),[40] 2D-$Mo_2C$ (17.91 mmol $g^{-1}$ $h^{-1}$ at 250 °C),[41] Fe-Pt/$CeO_2$ (434.7 mmol $g^{-1}$ $h^{-1}$ at 350 °C),[22] Pd/$TiO_2$ (216 mmol $g^{-1}$ $h^{-1}$ at 400 °C),[25] Pt/$CeO_2$ (172.8 mmol $g^{-1}$ $h^{-1}$ at 400 °C),[36] 2D-$Mo_2C$ (214.3 mmol $g^{-1}$ $h^{-1}$ at 430 °C).[41] **c** The stability test of 2D $Ni_1Ag_{0.02}O_1$ for RWGS at 275 °C for 76 hours. The RWGS test condition in **a, b, c** is: Catalyst's amount =5 mg, $CO_2$ flow rate = 30 mL $min^{-1}$, $H_2$ flow rate = 30 mL $min^{-1}$. **d** Apparent activation energy (Ea) of 2D $Ni_1Ag_{0.02}O_1$, Ni. The RWGS test condition of 2D $Ni_1Ag_{0.02}O_1$ in **d** is: Catalyst's amount =2 mg, $CO_2$ flow rate = 30 mL $min^{-1}$, $H_2$ flow rate = 30 mL $min^{-1}$.

**Asymmetric sites activated RWGS**

The mechanism of RWGS over 2D $Ni_1Ag_{0.02}O_1$ was investigated by Spin-polarized density functional theory (DFT).[42] Ag single atom boned on NiO(200) plane was used to represent 2D $Ni_1Ag_{0.02}O_1$ and Ni(111) plane was used as the reference sample (Supplementary Fig. 16). And, I*n-situ* DRIFTS spectroscopy was applied to identify the intermediates to infer reaction pathways (Fig. 3a). For the adsorbed $CO_2$, 2D $Ni_1Ag_{0.02}O_1$ showed the main peak at ~1597 $cm^{-1}$, faint peaks at 1338 $cm^{-1}$, 1238 $cm^{-1}$, assigned to bridged $CO_2$ (b-$CO_3$*), monodentate carbonate (m-$CO_3$*), $HCO_3$*, respectively.[43] Whereas, Ni only showed the monodentate carbonate adsorption Supplementary Fig. 17). Based on the adsorption results, we did the DFT calculation. For 2D $Ni_1Ag_{0.02}O_1$, the Bader charge of Ag and Ni in 2D $Ni_1Ag_{0.02}O_1$ was +1.18 and +0.12 |e|, respectively. It made Ni in 2D $Ni_1Ag_{0.02}O_1$ tend to adsorb electron rich oxygen ion of $CO_2$ and Ag single atom have strong interaction with electron deficient carbon ion in $CO_2$ (Supplementary Fig. 16). The asymmetric adsorption of $CO_2$ in 2D $Ni_1Ag_{0.02}O_1$ made C=O bond extend to 1.38 Å, in comparison with the 1.2 Å of C=O length in $CO_2$ adsorbed on Ni(111), and this elongated C=O bond was easily activated.[44] In comparison, the $CO_2$ was adsorbed on Ni(111) by Ni-C bond (Supplementary Fig. 16). At 150 °C of reaction temperature, the intermediate of COOH* (1750 $cm^{-1}$) appeared on 2D $Ni_1Ag_{0.02}O_1$,[45] which is the key intermediate in the reduction of $CO_2$ to CO, and the peak intensity of COOH* grows clearly with increasing reaction temperature (Fig. 3a). For the carboxyl route of RWGS shown in Fig. 3 b,c, the formation of the carboxyl group was the rate-determining step. The elongated O=C=O on 2D $Ni_1Ag_{0.02}O_1$ could break as O=C-O- to bond the H atom to form O=C-OH (1.03 eV for 2D $Ni_1Ag_{0.02}O_1$, 1.36 eV for Ni, Fig. 3b,c).[46] Fig. 3 d,e also show the RWGS process via formate route and the rate-determining step was the C=O bond breaking of formate. Due to the elongated C=O of $CO_2$, the C=O bond on 2D $Ni_1Ag_{0.02}O_1$ was easily split than that on Ni(111) (0.84 eV for 2D $Ni_1Ag_{0.02}O_1$, 1.46 eV for Ni, Fig. 3d,e). These results confirmed that 2D $Ni_1Ag_{0.02}O_1$ could activate $CO_2$ by the synergistic effect of Ag single atom and neighboured Ni site. In addition to the reaction energy barrier, we found that the desorption of CO was also the key for RWGS. In Ni(111), the CO was adsorbed by Ni site with 1.53 eV of desorption energy (Supplementary Fig. 18a, Fig. 3b).[47 48] Due to asymmetric adsorption of $CO_2$, the CO was adsorbed on the Ag single atom of 2D $Ni_1Ag_{0.02}O_1$ as shown in Supplementary Fig. 18b with desorption energy of 0.44 eV (Fig. 3 b,d). Ag species had weak adsorption for CO intrinsically,[49] thus leading to the low adsorption barrier of CO.

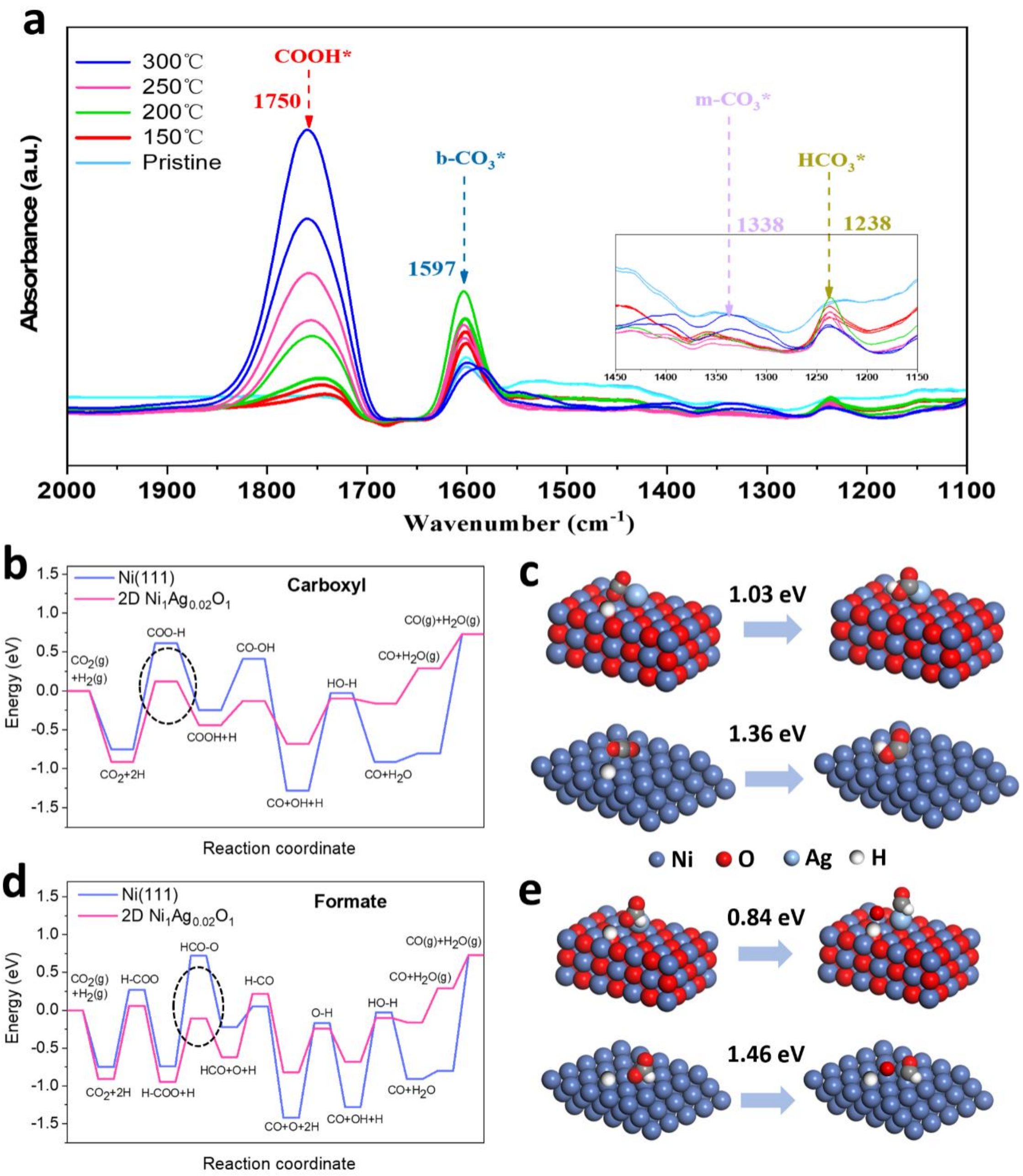

**Fig. 3** Ag-O-Ni synergistic effect and the reaction pathways for RWGS. **a** *In-situ* diffuse reflectance infrared Fourier transform spectra (DRIFTS) obtained for 2D $Ni_1Ag_{0.02}O_1$ under RWGS operating condition. **b** Energy profiles of the RWGS via carboxyl route on Ni(111) and 2D $Ni_1Ag_{0.02}O_1$. **c** The configurations of rate-determining intermediates of the carboxyl route on Ni(111) and 2D $Ni_1Ag_{0.02}O_1$. **d** Energy profiles of the RWGS via formate route on Ni(111) and 2D $Ni_1Ag_{0.02}O_1$. **e** The configurations of rate-

determining intermediates of the formate route on Ni(111) and 2D $Ni_1Ag_{0.02}O_1$.

**Weak sunlight driven photothermal RWGS**

Due to the excellent low-temperature performance, 2D $Ni_1Ag_{0.02}O_1$ was applied for weak sunlight driven photothermal RWGS.[17] 2D $Ni_1Ag_{0.02}O_1$ was placed into the TiC/Cu based device for photothermal RWGS (Supplementary Fig. 19).[50,51] Just under 0.1 kW $m^{-2}$ intensity of sunlight (0.1 sun) irradiation, the photothermal RWGS was clearly onset with the catalyst's irradiation temperature reaching 125 °C (Fig. 4a). When the irradiated sunlight intensity increased to 0.5 sun and 1 sun, the catalyst's temperature was promoted to 264 °C and 350 °C, respectively. Meanwhile, the corresponding photothermal RWGS CO generation rate soared to 650 mmol $g^{-1}$ $h^{-1}$ and 1065 mmol $g^{-1}$ $h^{-1}$, respectively. The CO selectivity of this system remained higher than 95% throughout the process (Supplementary Fig. 20). Fig. 4b and Supplementary Table 3 show the state of the art of photothermal RWGS. In previous reports, only when the light irradiation intensity exceeded 14 suns did CO production rate of photothermal RWGS surpass 100 mmol $g^{-1}$ $h^{-1}$. Therefore, the 1 sun-driven photothermal RWGS CO production rate (1065 mmol $g^{-1}$ $h^{-1}$) over 2D $Ni_1Ag_{0.02}O_1$ is not only the record performance under weak sunlight irradiation (1.01 mmol $g^{-1}$ $h^{-1}$ under 1 sun irradiation),[52] but also higher than the best value of advanced catalysts under concentrated sunlight irradiation (978.4 mmol $g^{-1}$ $h^{-1}$ under 28 suns irradiation).[20] Since the 2D $Ni_1Ag_{0.02}O_1$ can be scalably synthesized (Supplementary Fig. 21), we tested the performance of the photothermal system for RWGS. 30 g of granulated 2D $Ni_1Ag_{0.02}O_1$ was placed in the TiC/Cu based device with 0.036 $m^2$ of sunlight irradiation area. When the sunlight density was 0.4 sun, 4.88 L $h^{-1}$ of CO generation rate appeared through the 2D $Ni_1Ag_{0.02}O_1$ loaded in the TiC/Cu based device (Fig. 4c). With increasing the light intensity to 1 sun, the CO generation rate increased to 30.2 L $h^{-1}$ (Fig. 4c). Based on the experimental data, the solar to chemical energy conversion efficiency (STC) of photothermal RWGS reached 39.1% under 1 sun irradiation (Fig. 4c). Therefore, the 2D $Ni_1Ag_{0.02}O_1$ made photothermal catalysis open a new pathway for achieving efficient weak solar driven $CO_2$ hydrogenation.

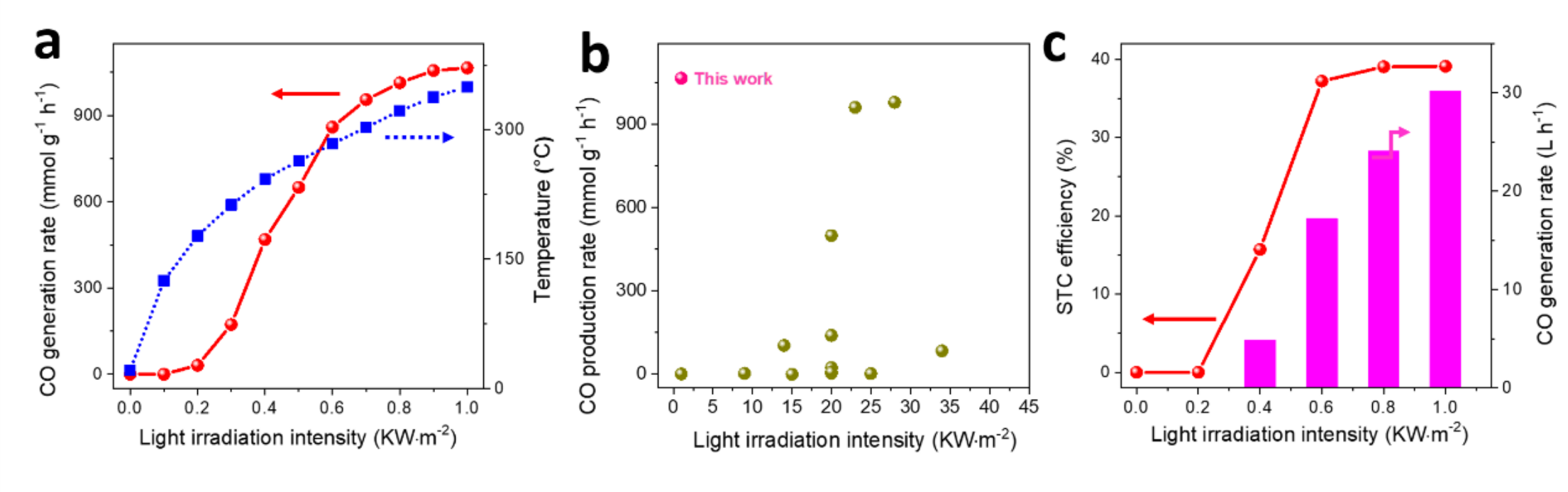

**Fig. 4** Weak sunlight driven photothermal RWGS. **a** The sunlight irradiation temperature and CO production rate of 2D $Ni_1Ag_{0.02}O_1$ combined with TiC/Cu based device. **b** The photothermal RWGS performance of reported state of the art of catalysts, e. g., $BiO_x/CeO_2$ (1.01 mmol $g^{-1}$ $h^{-1}$, 1 sun),[52] Ni@p-$SiO_2$ (978.4 mmol $g^{-1}$ $h^{-1}$, 28 suns),[20] $Ni_{12}P_5/SiO_2$ (960 mmol $g^{-1}$ $h^{-1}$, 23 suns),[21] Co-PS@$SiO_2$ (498.9 mmol $g^{-1}$ $h^{-1}$, 20 suns),[53] CF-$Cu_2O$ (139.6 mmol $g^{-1}$ $h^{-1}$, 20 suns),[54] 2D $In_2O_{3-x}$ (103.21 mmol $g^{-1}$ $h^{-1}$, 14 suns),[55] Ru/$Mo_2TiC_2$ (84 mmol $g^{-1}$ $h^{-1}$, 34 suns),[56] $In_2O_{3-x}$ (23.8 mmol $g^{-1}$ $h^{-1}$, 20 suns),[57] $Bi_xIn_{2-x}O_3$ (8 mmol $g^{-1}$ $h^{-1}$, 20 suns),[58] Pd@$H_yWO_{3-x}$ (3 mmol $g^{-1}$ $h^{-1}$, 20 suns),[59] Rh-$In_2O_{3-x}(OH)_y$ (2.4 mmol $g^{-1}$ $h^{-1}$, 9 suns),[38] Pd/$Nb_2O_5$ (1.8 mmol $g^{-1}$ $h^{-1}$, 25 suns),[60]

Pd@SiNS (0.01 mmol $g^{-1}$ $h^{-1}$, 15 suns).[61] The photothermal RWGS test condition in **a, b** is: catalyst's amount =10 mg, $CO_2$ flow rate = 30 mL $min^{-1}$, $H_2$ flow rate = 30 mL $min^{-1}$. **c** The CO production rate and STC efficiency of scalable photothermal RWGS. The photothermal RWGS test condition in **c** is: catalyst's amount =30 g, $CO_2$ flow rate = 200 L $h^{-1}$, $H_2$ flow rate = 200 L $h^{-1}$.

**The application on outdoor artificial photosynthesis**

To realize the application of artificial photosynthesis, artificial photosynthesis system should have the characteristics of outdoor sunlight driven operation, large working area and high utilization rate of raw materials. As shown in Fig. 5a, an outdoor artificial photosynthesis system with 103 $m^2$ scale was built by using photovoltaic-electrocatalysis (94 $m^2$ of irradiation area) to split water ($2H_2O \rightarrow 2H_2 + O_2$)[62] and using 2D $Ni_1Ag_{0.02}O_1$ assisted photothermal catalysis (9 $m^2$ of irradiation area) to hydrogenate $CO_2$ ($CO_2 + H_2 \rightarrow CO + H_2O$).[45,63,64] In this system, the hydrogen generated from photovoltaic water electrolysis[65] was directly injected into the photothermal system for $CO_2$ hydrogenation driven by sunlight.[18,57,66] On April 5, 2023, an outdoor artificial photosynthetic system for CO production was tested in Baoding City, Hebei Province, China, with an ambient temperature ranging from 7 to 19 °C during the daytime. The sunlight irradiation area of the system was 103 $m^2$, and 2 kg of 2D $Ni_1Ag_{0.02}O_1$ was employed as the catalyst for photothermal RWGS. The outdoor sunlight intensity varied between 0.37 and 0.76 kW $m^{-2}$ (Supplementary Fig. 22). As shown in Fig. 5b, when photovoltaic electrocatalysis generated $H_2$ at 8:30 AM (Supplementary Fig. 23), the photothermal system synchronously converted $H_2$ and $CO_2$ as CO, and the CO generation rate at 9:00 AM reached 1.83 $m^3\ h^{-1}$. On April 5, 2023, the CO generation rate in the outdoor artificial photosynthetic system increased to a peak value at 3.2 $m^3\ h^{-1}$ at 12:00 PM, and then gradually decreased to 1.52 $m^3\ h^{-1}$ at 16:00 PM. Although the solar intensity and ambient temperature are lower, the outdoor system STC for CO products can still be >16% throughout the operating period (Fig. 5c, detailed calculation seen in Methods). The daily total CO output of the system was ~18.70 $m^3$. In the whole process, more than 90% of green $H_2$ was involved in the RWGS (Supplementary Fig. 24), which proved that the system can not only efficiently chemically store green $H_2$ generated by photovoltaic electrocatalysis, but also fix $CO_2$ on a large scale. Fig. 5d listed the STC values of reported advanced artificial photosynthetic systems with > 1 $cm^2$ of irradiation area. Firstly, the size of our outdoor artificial photosynthetic system is 103 $m^2$, 15.8 times larger than the reported large-scale artificial photosynthesis (6.5 $m^2$),[15] demonstrating its potential for mass production. Although used silicon solar cell, the STC of our outdoor demostration for $CO_2$ reduction as CO is ~16.4%, which is not only 4.3 times higher than the STC of large-scale artificial photosynthesis (3.8%),[15] but also 1.3 times higher than that of reported advanced artificial photosynthetic systems (12.7%) by using triple-junction solar cells.[16] Consequently, the outdoor artificial photosynthetic system outperforms existing systems in terms of both scalability and conversion efficiency, making it a promising candidate for practical applications in $CO_2$ reduction and artificial photosynthesis.

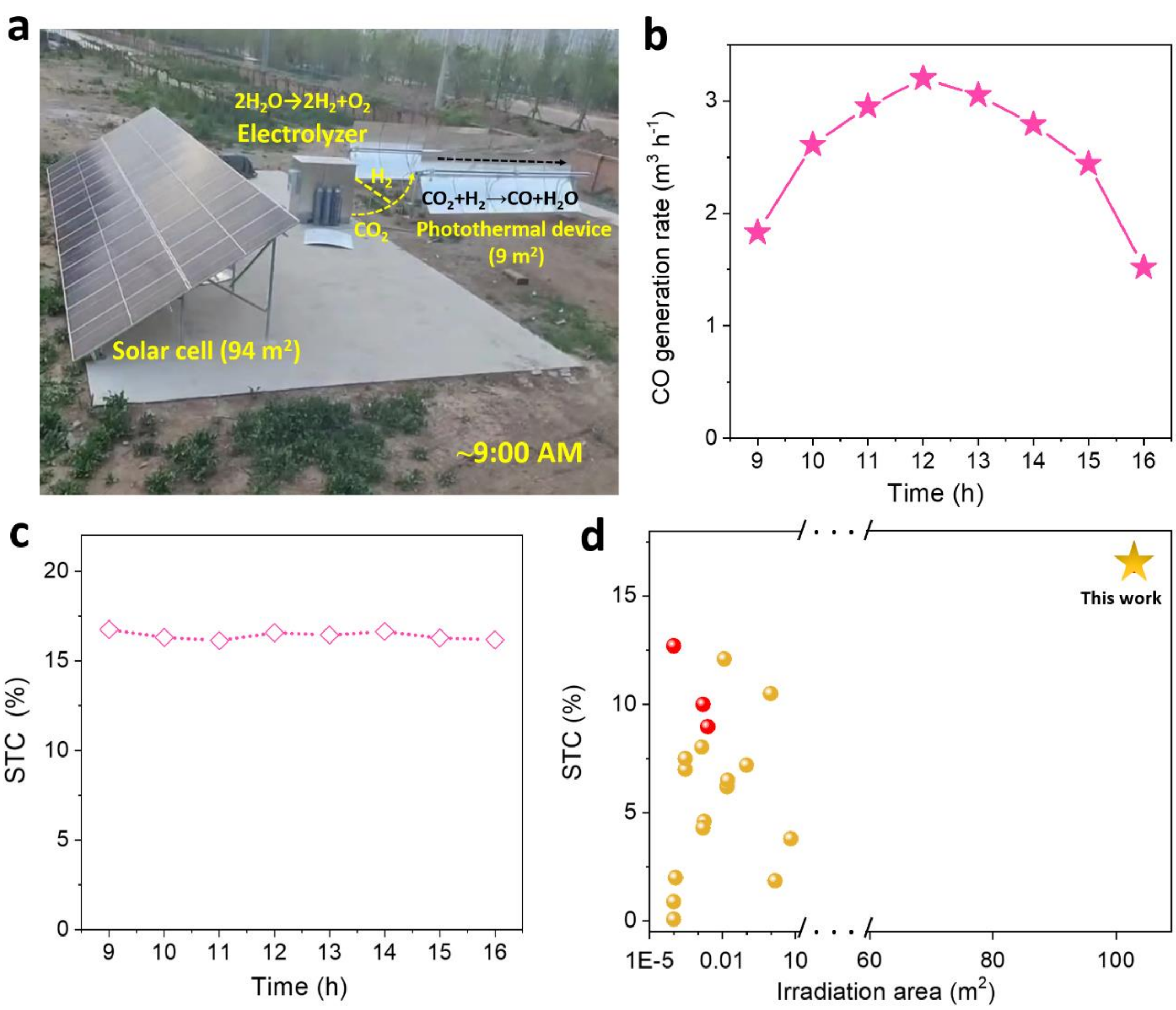

**Fig. 5** The outdoor performance of artificial photosynthetic system. **a** The photograph of new artificial photosynthetic demonstration in Hebei University. **b, c** The CO production rate and STC of new artificial photosynthetic demonstration under ambient sunlight irradiation, on April 05, 2023, in Baoding City, China. **d** Comparison of the solar driven $CO_2$ reduction systems of this work and the state of the art of artificial photosynthetic systems. The red spot represents that applied solar cell are triple-junction solar cell (28%-33% efficiency).[67] The STC and irradiation area of reported state of the art of artificial photosynthetic systems are as follws: (1.86%, 1.47 $m^2$),[68] (3.8%, 6.5 $m^2$),[15] (4.3%, 16 $cm^2$),[69] (4.6%, 18 $cm^2$),[70] (6.2%, 156 $cm^2$),[71] (6.5%, 165

cm$^2$),[72] (7.2%, 987 cm$^2$),[5] (8.03%, 14 cm$^2$),[73] (8.97%, 25 cm$^2$),[74] (10.5%, 0.96 m$^2$),[6] (12.1%, 120 cm$^2$),[75] (2%, 1.18 cm$^2$),[76] (7%, 3 cm$^2$),[77] (10%, 16 cm$^2$),[78] (7.5%, 3.198 cm$^2$),[79] (0.08%, 1 cm$^2$),[80] (0.9%, 1 cm$^2$),[81] (12.7%, 1 cm$^2$).[16]

## Conclusion

In this work, a starch assisted templated method was developed to synthesize Ag single atoms supported on NiO nanosheets (2D $Ni_1Ag_{0.02}O_1$). The resulting catalyst demonstrated excellent RWGS performance, with a CO production rate of 1434 mmol $g^{-1}$ $h^{-1}$ at 300 ℃. *In-situ* DRIFTS and theoretical calculation indicated that the $CO_2$ was asymmetrically adsorbed on Ag-O-Ni sites of 2D $Ni_1Ag_{0.02}O_1$, leading to the elongate C=O bond (1.38 Å). This enabled the active hydrogenation of $CO_2$ while also resulting in weak CO adsorption (0.44 eV). With the assistance of TiC/Cu based device, the 2D $Ni_1Ag_{0.02}O_1$ catalyst exhibited remarkable photothermal RWGS performance under 1-sun irradiation, with a CO production rate of 1065 mmol $g^{-1}$ $h^{-1}$. Moreover, 94 $m^2$ of silicon solar cell was used to drive the electrolyzer for photovoltaic electrolytic water splitting as $O_2$ and $H_2$, then, the generated $H_2$ and $CO_2$ were injected into the TiC/Cu based device (9 $m^2$) loaded with 2D $Ni_1Ag_{0.02}O_1$ to carry out photothermal $CO_2$ hydrogenation. This system demonstrated a CO production of 18.70 $m^3$ per day and the solar to chemical energy conversion efficiency of >16% all day-time, under ambient solar irradiation. This work revealed that Ag single atom modification provided a new route to construct low-temperature $CO_2$ hydrogenation catalysts without platinum-group-metals. Based on the low-temperature $CO_2$ hydrogenation activity, 2D $Ni_1Ag_{0.02}O_1$ could drive ambient sunlight driven photothermal RWGS and new artificial photosynthesis to realize natural sunlight-driven carbon neutralization. This breakthrough has significant implications for the development of sustainable energy solutions and environmental protection, offering a viable alternative to artificial photosynthesis and traditional catalysts.

# Methods

## Chemicals for catalysts

Commercial silver nitrate ($AgNO_3$), nickel nitrate hydrate ($Ni(NO_3)_2 \cdot 6H_2O$), water-soluble starch, Ag nanoparticles were bought from Sinopharm Co., Ltd. The chemicals were all used without any further treatment.

## 2D $Ni_1Ag_{0.02}O_1$

Firstly, 10g water-soluble starch, 12 mg $AgNO_3$ and 972 mg $Ni(NO_3)_2$ $6H_2O$ were dissolved into 400 mL of water. After 0.5 hours of stirring, the uniform solution was dripped into liquid nitrogen to make it freeze into ice quickly and it was freeze-dried for 48 hours to remove $H_2O$. The dried product was calcined in a muffle furnace at 425 ℃ (heating rate 1 ℃ $min^{-1}$) for 17 hours, and the obtained was named 2D $Ni_1Ag_{0.02}O_1$.

## Ni

Firstly, 10g water-soluble starch, 972 mg $Ni(NO_3)_2$ $6H_2O$ were dissolved into 400 mL of water. After 0.5 hours of stirring, the uniform solution was dripped into liquid nitrogen to make it freeze into ice quickly and it was freeze-dried for 48 hours to remove $H_2O$. The dried product was calcined in a muffle furnace at 425 ℃ (heating rate 1 ℃ $min^{-1}$) for 17 hours, and then the annealed sample was reduced by $H_2$ at 400 ℃ for 5 hours to achieve the sample.

**Photothermal RWGS**

The photothermal RWGS was tested as follows: 10 mg of 2D $Ni_1Ag_{0.02}O_1$ was loaded intoTiC/Cu based device (0.036 $m^2$), and irradiated by a light source (DL-3000). In this test, the flow of feed gas was the mixture of 30 mL $min^{-1}$ of $CO_2$ and 30 mL $min^{-1}$ of $H_2$. The reaction products were tested by gas chromatography (GC) 7890A equipped with FID and TCD detectors.

For the scalable test, the amount of 2D $Ni_1Ag_{0.02}O_1$ was increased to 30 g, the flow of feed gas was the mixture of 200 L $h^{-1}$ of $CO_2$ and 200 L $h^{-1}$ of $H_2$.

**Enthalpy change energy of chemicals**

The enthalpy change energy of $CO_2$ (g), CO (g), $H_2$ (g), $O_2$ (g), $H_2O$ (g), $H_2O$ (l) was -393.505, -110.541, 0, 0, -241.818, -285.830 kJ $mol^{-1}$, respectively.

The (g) and (l) indicated the gas state and liquid state, respectively.

**Solar to chemical energy conversion efficiency (STC) of photothermal RWGS**

The STC of photothermal RWGS demonstration was calculated as follows:

$$STC= (\Delta H*\varepsilon/24.5)/(I*S*3600) \qquad (1)$$

ΔH was the reaction Enthalpy change energy ($CO_2$ (g) + $H_2$ (g) → CO (g) + $H_2O$ (g), RWGS, ΔH= 41.15 kJ $mol^{-1}$), ε (L $h^{-1}$) was the CO generation amount per hour detected by a flowmeter, I was the light intensity (kW $m^{-2}$), S was the irradiated area of demonstration (0.036 $m^2$). The ε irradiated by 0.4 sun, 0.6 sun, 0.8 sun, 1 sun was 4.88 L $h^{-1}$, 17.22 L $h^{-1}$, 24.10 L $h^{-1}$, 30.20 L $h^{-1}$, respectively, corresponding to 15.7%, 37.2%, 39.0%, 39.1% of STC, severally.

**Outdoor artificial photosynthetic system**

The outdoor artificial photosynthetic system consisted of two components. One component was the photovoltaic electrolysis system, in which the solar cell (TSM-DE19) with 94 $m^2$ of solar irradiation area was used to power electrolytic reactor. The mixture of 30 kg KOH and 100 L deionized water was used as the electrolyte. The other component was the photothermal system, in which a TiC/Cu based photothermal device was provided by Hebei scientist research experimental and equipment trade Co., Ltd. with the size of 4 cm in diameter and 6 m inlength, equipped with a reflector of 6 m inlength and 1.5 m in width. For the production of CO, the catalyst used in TiC/Cu based photothermal device was 2 kg 2D $Ni_1Ag_{0.02}O_1$. For CO production in photothermal system, the $CO_2/H_2$ ratio was 12. The data were collected by FID and TCD.

**The STC of outdoor artificial photosynthetic system**

The STC of the outdoor artificial photosynthetic system for converting $CO_2$ into CO was calculated as follows:

$$STC= (\Delta H*\varepsilon)/(I*S*3.6*22.4) \qquad (2)$$

ΔH was the reaction Enthalpy change energy ($H_2O$ (l) + $CO_2$ (g) → CO (g) + 1/2 $O_2$ (g) + $H_2O$ (g), ΔH= 326.9754 kJ $mol^{-1}$), ε ($m^3$) was the CO generation amount per hour detected by a flowmeter, I was the outdoor solar intensity (kW $m^{-2}$), and the I at 9:00 AM, 10:00 AM, 11:00 AM, 12:00 AM, 13:00 PM, 14:00 PM, 15:00 PM, 16:00 PM was 0.43 kW $m^{-2}$, 0.63 kW $m^{-2}$, 0.72 kW $m^{-2}$, 0.76 kW $m^{-2}$, 0.73 kW $m^{-2}$, 0.66 kW $m^{-2}$, 0.59 kW $m^{-2}$, 0.37 kW $m^{-2}$, respectively. S was the total irradiated area of 103 $m^2$. The ε at 9:00 AM, 10:00 AM, 11:00 AM, 12:00 AM, 13:00 PM, 14:00 PM, 15:00 PM, 16:00 PM was 1.83 $m^3$ $h^{-1}$, 2.61 $m^3$ $h^{-1}$, 2.95 $m^3$ $h^{-1}$, 3.20 $m^3$ $h^{-1}$, 3.05 $m^3$ $h^{-1}$, 2.79 $m^3$ $h^{-1}$, 2.44 $m^3$ $h^{-1}$, 1.52 $m^3$ $h^{-1}$, respectively, corresponding to 16.75%, 16.30%, 16.13%, 16.57%, 16.44%, 16.64%, 16.28%, 16.17% of STC, severally.

## Data availability

The data generated in this study are provided in the main text and Supplementary information. Extra data are available from the corresponding author upon reasonable request. Source data are provided with this paper.

## Acknowledgements

This work is supported by the Natural Science Foundation of Hebei Province (Grant Nos. B2022201090, B2021201074, B2021201034, F2021203097), Interdisciplinary Research Program of Natural Science of Hebei University (Grant Nos. 521100311, DXK202109), Hebei University (050001-521100302025, 050001-513300201004), the Knowledge Innovation Program of the Chinese Academy of Sciences. Thank you for the TEM technical support provided by the Microanalysis Center, College of Physics Science and Technology, Hebei University.

## Author contributions

Y. L., Q. M. conceived the project and contributed to the design of the experiments and analysis of the data. Q. W. and D. Y. performed the catalysts preparation and characterizations. H. W., B. L., and J. W. performed the photothermal devices' characterizations. F. M., X. S., L. G. conducted the SEM and TEM examinations. S. W. provided the optical advice. Y. L. and Q. M. wrote the paper. All the authors discussed the results and commented on the manuscript.

## Competing interests

The Authors declare no competing interests.

## Additional information

**Supplementary Information** accompanies this paper.

**Correspondence** and requests for materials should be addressed to Yaguang Li or Qingbo Meng.